# Decreasing the Zeeman shift uncertainty by real-time magnetic field compensation

Wenli Wang, Richang Dong, Rong Wei, Tingting Chen, Qian Wang, and Yuzhu Wang

*Abstract*—We present a dynamic compensation method to compress the spatial fluctuation of the static magnetic field (C-field) that provides a quantization axis in the atomic fountain clock. The coil current of the C-field is point-by-point modulated in accordance with the atoms probing the magnetic field along the flight trajectory. A homogeneous field with a 0.2 nT inhomogeneity is produced compared to a 5 nT under the static magnetic field with a constant current during the Ramsey interrogation. The corresponding uncertainty associated with the second-order Zeeman shift that we calculate is improved by one order of magnitude. The technique provides an alternative method to improve the uniformity of the magnetic field, particularly for large-scale equipment that is difficult to construct with an effective magnetic shielding. Our method is simple, robust, and essentially important in frequency evaluations concerning the dominant uncertainty contribution due to the quadratic Zeeman shift.

*Index Terms*—Magnetic variables control, magnetic field measurement, magnetic field effects, performance evaluation, atomic clocks

## I. Introduction

THE output of a primary frequency standard based on atomic fountain gives the definition of the second in the International System of Units (SI). Currently, relative uncertainties on the order of $10^{-16}$ are obtained with primary frequency standards.[1] To achieve superior performance, evaluations of the systematic frequency shifts and the corresponding uncertainty become the most important work, both in atom clocks[2-5] and atom interferometers.[6-8] When atoms interact with the microwave field in a fountain clock or with the light field in an interferometer, a uniform and stable magnetic field (C-field) is required to define the quantization axis. Normally, the existence of the magnetic field produces the largest frequency biases, and the noise introduced by the associated homogenous or inhomogeneous broadening mechanism leads to one of the largest uncertainty contributions to the output frequency of the standard. Therefore, the suppression of the C-field inhomogeneity plays a key role in accurately evaluating the Zeeman shift and improving the clock's performance.

For most atomic fountain clocks (AFCs), the stray magnetic field along the atomic trajectories is blocked by enclosing the solenoid coil within multiple-layers of magnetic shields. Meanwhile, several techniques such as microwave radiation [9], polarization rotation [10], Larmor spin precession [11], stimulated Raman transition (SRT) [12-15], and light-mediated quantum interference effect [16] have been applied for the precise measurement of the strength of the magnetic field generated by a solenoid. For example, at the National Institute of Standards and Technology (NIST), four layers of cylindrical magnetic shields enclose the drift tube and the Ramsey cavity. The magnetic field strength shows an inhomogeneity of less than 6 nT. The calculated second-order Zeeman shift at the C-field strengths typically used in NIST-F1 is about 400 µHz [4]. In the rubidium fountain clock that we have constructed, a four-layered magnetic shielding structure has been built, with a static magnetic field homogeneity of 5 nT. Such a value corresponds to an uncertainty of $4\times10^{-16}$ along the atomic trajectory, in the quadratic Zeeman shift.

In our fountain clock, considering the inhomogeneity of the static magnetic field distribution that provides a quantization axis, we utilize a dynamic compensation method to compress the spatial fluctuation of the biasing magnetic field along the atomic trajectory, combined with the SRT method for the accurate mapping of the magnetic field vector [17]. In the moving reference frame of the cold atoms, based on the dependence of the magnitude of the C-field on the coil current, a real-time feedback procedure for the local magnetic field is described. Compared with the static method, the inhomogeneity of the magnetic field probed by the atoms is reduced by one order of magnitude. We also experimentally analyze the phenomenon concerned with the single-point modulation, and several steps using a point-by-point compensation method are repeatedly applied to obtain a reduced systematic uncertainty. Finally, an overview of the temporal stability of the local magnetic field is briefly discussed. Our method is effective and provides a promising alternative approach for large-scale devices that are difficult to

Manuscript submitted April 21, 2017. This work was supported by the National Natural Science Foundation of China under Grant Nos. 11404353, 91336105 and 61275204, and the Strategic Priority Research Program of the Chinese Academy of Sciences，Grant No. 8 (XDB21030800).

W. L. Wang, R. Wei and Y. Z. Wang are with the Key Laboratory for Quantum Optics, Shanghai Institute of Optics and Fine Mechanics, Chinese Academy of Sciences, Shanghai 201800, China.
R. C. Dong, T. T. Chen and Q. Wang are with the Key Laboratory for Quantum Optics, Shanghai Institute of Optics and Fine Mechanics, Chinese Academy of Sciences, Shanghai 201800, China. and with the University of Chinese Academy of Sciences, Beijing 100049, China.
(email: weirong@siom.ac.cn)

construct with magnetic shielding [8].

## II. EXPERIMENTAL SETUP

A schematic diagram of our rubidium vacuum part is illustrated in Fig. 1(a). The fountain described here uses a (1, 1, 1) configuration for the laser cooling and launching atoms where a folded optical path is used [18]. A state-selection microwave cavity, a detection zone, and a Ramsey cavity are mounted above the cooling zone. A detailed description of the atomic fountain is given our previous work [9]. In this work, both the state selection cavity and Raman cavity are not excited. The sample of cold atoms (approximately 2 μK) is distributed evenly in each $|F=2, m_F\rangle$ sublevel and launched vertically upwards through the detection region. After passing through the excitation cavity, the atoms enter the C-field region and begin a free fall under the influence of gravity, successively re-entering the detection region. In our rubidium fountain clock, a small static magnetic field inside the innermost shield is applied by a precision-wound solenoid providing the quantization axis. This field will produce a second-order Zeeman shift of the clock transition frequency, which can be calculated from the measured value of the magnetic field. To improve the axial homogeneity of the C-field region, two pairs of additional coaxial short-length coils are positioned close to each end of the main coil. Four layers of magnetic shields made of mu-metal enclose the Ramsey cavity and atomic flight region, which reduce the influence of the Earth's magnetic field to less than 5 nT.

We used the method of SRT to map the magnetic strength along the interrogation zone of the AFC. As shown in Fig. 1(b), the carrier and the +1st order sideband modulated by a fiber electro-optic modulator (FEOM) are used as the two Raman beams and injected from the bottom window of the vacuum chamber. The radio frequency (RF) signal is supplied by a signal generator coupled to a Hydrogen maser (VCH-1003A), and the frequency of the RF signal is swept to excite the corresponding two-photon transitions. To produce a large detuning from the excited state F'=2, the input laser was red shifted to 200 MHz by a double-pass optical path using an acoustic optical modulator (AOM).

The scheme of the used transition energy levels of $^{87}$Rb for SRT with a large one-photon detuning $\Delta_s$ is shown in Fig. 1(c). Since there is an angle between the polarization direction of the Raman beams and the direction of the magnetic field, as described in our previous work, all possible electric-dipole transitions with $\Delta m=0, \pm1, \pm2$ will occur.

By controlling the location of the atoms and the interaction time with the Raman light fields, the relative atom populations $N_2$ and $N_1$ in the F=2 and F=1 hyperfine levels, respectively, are measured by the probe pulses, and used to calculate the normalized transition probability as $P=N_2/(N_1+N_2)$. The frequency detuning of the Raman beams is increased progressively by small increments, and the full map of the magnetic field strength is constructed from a number of different measurements at discrete locations along the atomic flight region. In this work, we used the Raman spectroscopy of

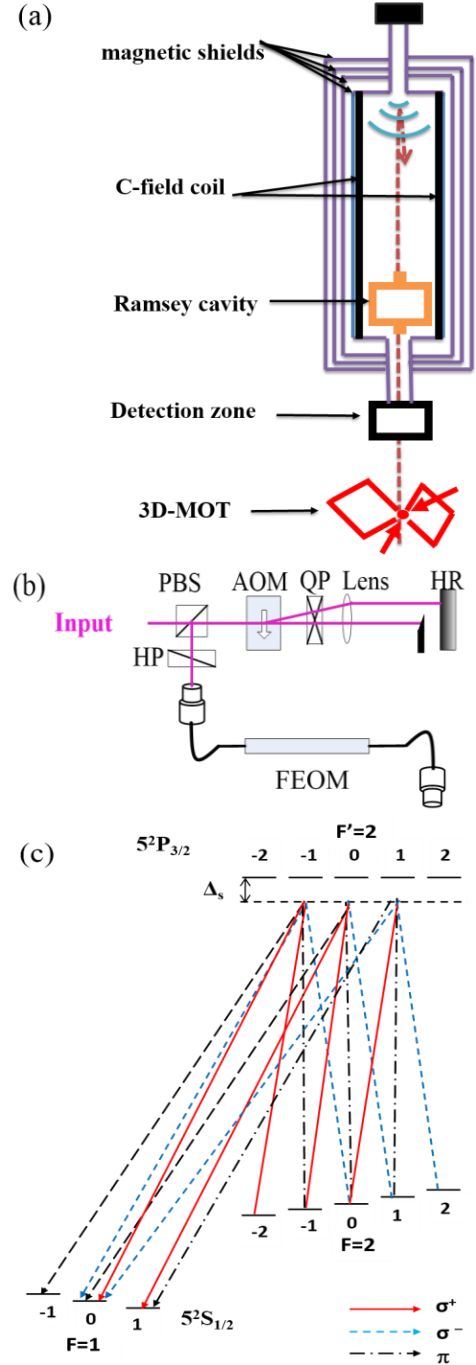

Fig. 1. (a) Mechanical drawing of the $^{87}$Rb-fountain physics package. In this work, the state selection cavity and Raman cavity are turned off. (b) The main laser set-up used for generation of the Raman beams. The Raman beams are injected from the bottom window of the magneto-optical trap (MOT). HP—half-wave plate, QP—quarter-wave plate, PBS—polarization-beam-splitter, AOM—acoustic optical modulator, FEOM—fiber electro-optic modulator. (c) Levels for stimulated Raman transitions with a large one-photon detuning $\Delta_s$ and two-photon detuning $\delta$ under the condition that the $^{87}$Rb atoms are distributed at $|F=2, m_F\rangle$ sublevels. Seven Λ systems are allowed by the two-photon electric-dipole selection rules, $\Delta m=0, \pm1, \pm2$.

the $|1, -1\rangle$ ($|F=1, m_F=-1\rangle$)→$|2, -1\rangle$ and $|1, 1\rangle$→$|2, 1\rangle$ field sensitive transitions to measure the complete magnetic field distribution. The magnetic field strength is proportional to the current in the solenoid, which is obtained from a DC voltage

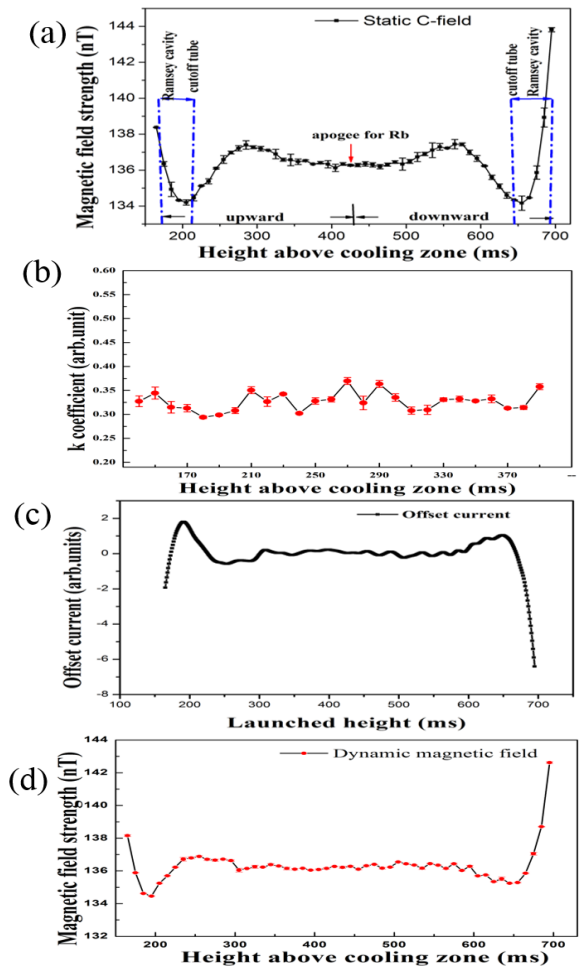

Fig. 2. (a) Mapping of the static magnetic field along the atomic trajectory above the cooling zone. The entire height region contains the upward and downward flight processes of the atoms. The red arrow denotes the apogee for the Rb atoms. The positions of the Ramsey cavity and the cutoff tubes are indicated by the dotted box. (b) The tracing of the scale coefficients between the magnitude of the magnetic field and the solenoid current, along the atomic flight region. (c) The offset current of the solenoid as a function of the launch height. (d) The dynamic magnetic field distribution for the first step using our TDTC method. The error bars indicate the systematic uncertainty of our measurement.

source controlled by a digital-to-analog converter (DAC), in series with a precision resistor.

### III. EXPERIMENTAL METHOD

Figure 2(a) shows the static distribution of the magnetic field inside the inner shield during the atomic ballistic flight, with the coil current fixed at 0.43 mA. The measured value of the magnetic field $B(z)$ averaged over the ballistic time of flight is expressed as $<B(z)>$, where z is the launch height determined by the relation $z=v_0 t-1/2gt^2$, with the value of the launch velocity $v_0=4.188$ m/s and $<B(z)>= 136.43$ nT in our experiment. It shows an inhomogeneity of 5 nT. The inhomogeneity of the static magnetic field depends on the field leakage from the holes in the shielding, magnetic field coils, and the effect of the highly permeable vacuum feedthroughs on the Ramsey cavity.

In the moving reference frame of the cold atoms, the local magnetic field can be written as $B(z, I)$, which is determined by both the atomic position z and the current in the solenoid I. Experimentally, we measured the scale coefficient k between the magnitude of the magnetic field and the solenoid current, as shown in Fig. 2(b). It varies along the atomic ballistic flight region, and the error bars indicate the systematic uncertainty of our measurement.

In Fig. 2(c), we calculate the offset value of the solenoid current $\delta I$, expressed as $\delta I=\delta B/k$, where $\delta B$ is the magnetic field offset from the expected value of the magnetic field strength $B_0$, and in our case $B_0=136.05$ nT. In theory, by the real-time compensation of the offset current of the solenoid, it enables a further fine-tuning of the magnetic field along the atomic trajectory and realizes a local magnetic field probed by the atoms, as homogeneous as possible. Finally, by the real-time tracing of the scale coefficients along the atomic flight region, a data array for the compensation of the offset current is assigned and then recorded by a computer, which also controls the execution of all triggering pulses for our fountain operation.

The preliminary results of the magnetic field using our trajectory dynamic tracing compensation technique (TDTC) are plotted in Fig. 2(d). The magnetic field inhomogeneity in the interaction region is actively improved compared with the static method in Fig. 2 (a). However, there is still a slight deviation from the expected value. We concluded that the real-time compensation process along the atomic flight region was not realized. We measured the temporal response of the coil current under the condition that the inductance L=40 μH, resistance R=10 kΩ, and the time constant t=L/R=4 ns. Therefore, we excluded the cause induced by time delay of the coil inductance. We initially assumed that it may have been caused by the eddy effects in the shielding system.

### IV. RESULTS AND DISCUSSION

We experimentally studied the dynamic response to the real time compensation process by means of a single-point modulation. At the location of the launch height of 420 ms, we alternated the coil current by 10 μA, denoted by the blue arrow as shown in Fig. 3(a). It is important to note that, the influence on the several subsequent points was observed because of the 20 ms extension in our dynamic feedback process. To solve this problem, we applied the point-by-point method backwards for correcting the magnetic field. The standard deviations associated with each point include the resonance-curve center uncertainty and the positional uncertainty. After a number of repeated operations using our method, the full map of the magnetic field was generated, denoted by the red circle line as shown in Fig. 3(b), with a higher measurement precision compared with that of the static result (black square line). The inset shows the magnified red-line region. It shows an inhomogeneity of 0.2 nT in the dynamic magnetic field, and the error bars show the systematic uncertainty introduced due to our measurement, which also illustrates that our optimized results are limited by the measurement precision.

Besides the spatial inhomogeneity, the instability of the magnetic field accounting for temporal fluctuations creates a larger uncertainty in the correction value. We evaluated the temporal instability of the magnetic field measured by the SRT spectroscopy performed at a single point [2], as shown in Fig. 4. Black and red lines display the temporal instability of the magnetic field for a comparison, expressed by the total Allan deviation $\sigma_y(\tau)$. In our method, the total Allan deviation is the quadratic sum of the two contributions— the spatial distribution $B_z$ and the temporal fluctuation $B_t$ associated with the parameter I—expressed as $\sigma_y(\tau)=\sqrt{(\sigma_z^2+\sigma_t^2)}$. For averaging times around $8\times10^4$ s, the temporal instability at the nominal launch height (0.881 m) is $2.5\times10^{-5}$. We noticed that the temporal fluctuations in the magnetic field using our method did not deteriorate despite the additionally introduced variable I, which was the result we expected.

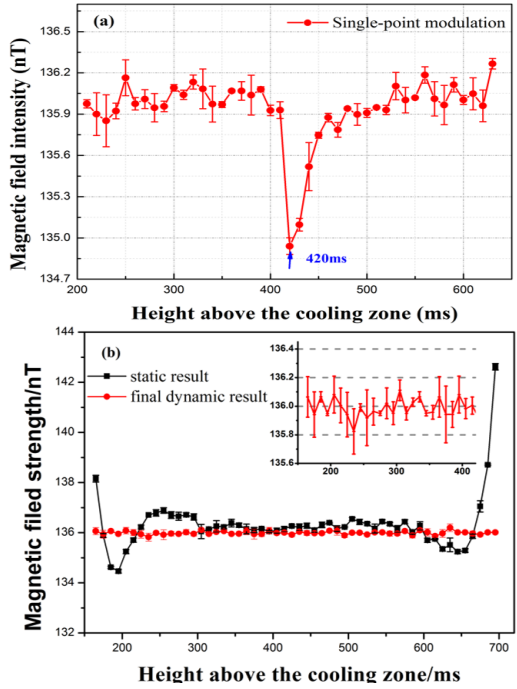

Fig. 3. (a)The dynamic response result of the corrected magnetic field strength above the cooling zone, where the voltage of the solenoid is varied by 10 μA at the single point position for a tossing time delay of 420 ms, denoted by the blue arrow. (b) Mapping of the magnetic field strength via several repeated steps of the dynamic compensation process. The black curve indicates the static results, while the red curve indicates the final results. The inset shows the magnified red-line region.

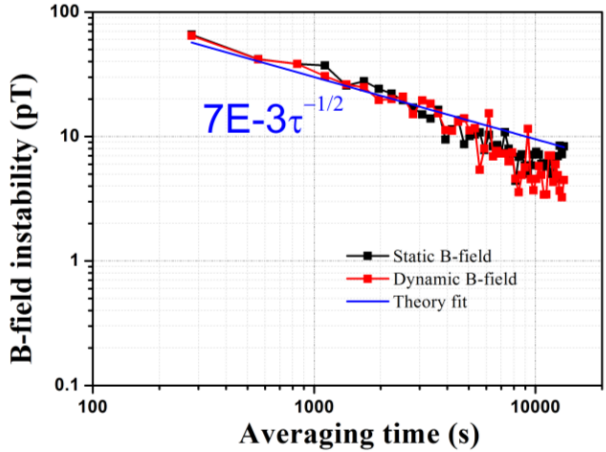

Fig. 4. Total Allan deviation of the temporal instability of the magnetic field for the fixed launch height of 420 ms. The black line represents the static method, while the red line represents our dynamic method. The blue line is the theoretically fitted curve.

In our experiment, the interrogation time of the Raman pulse was 10 ms, and the period of our measurement was 280 s. Subsequently, by increasing the spatial resolution of our magnetic field measurement, the long-term stability would be further improved. The blue line is the theoretically fitted curve, which demonstrates a typical long-term instability of $7E-3\tau^{-1/2}$, manifesting the characteristics of white noise. The contribution of the inhomogeneous broadening mechanism due to the gradient field also causes a frequency bias, and a related evaluation of the uncertainty contribution is in progress currently to clarify its origin and characteristics further. Compared with the static method, the fluctuation of the magnetic field distribution was improved by one order of magnitude, leading to an uncertainty of $4\times10^{-17}$ associated with the second-order Zeeman shift, directly evaluated by calculating the time average of the square of the magnetic field over the atomic trajectory, and led to perfect homogeneity of the C-field.

## V. CONCLUSIONS

In conclusion, we propose a dynamical compensation technique for compressing the C-field homogeneity in our $^{87}$Rb atomic fountain clock. In the moving reference frame of the cold atoms, an inhomogeneity of 0.2 nT for the local magnetic field is realized, with a reduced uncertainty contribution due to the second order Zeeman shift. The technique provides an alternative method to improve the magnetic field uniformity, particularly for large-scale equipment that is difficult to configure with magnetic shielding. Our work can be widely applied to other precision measurement fields, such as the atom gravimeter [19], atomic gyroscope [20,21], and fundamental physical experiments [22].

**Wenli Wang** received the Ph. D. degree from East China Normal University, Shanghai, China.
  She is an assistant research fellow at the Shanghai Institute of Optics and fine Mechanics (SIOM), Chinese academy of sciences in 2012. Her research work is principally in the area of fundamental physical sciences and technology in atomic clocks, especially the integrating sphere atomic clock and rubidium fountain clock.

**Rong Wei** was born in Taiyuan, Shanxi, China. He received his Ph. D. degree from Shanghai Institute of Optics and fine Mechanics (SIOM), Chinese academy of sciences in 2003. Since then on, he has been a researcher in SIOM.
  His research focused on atomic fountain clock and improvement of principle and method of atomic clock and clock group, which is based on two devices of a rubidium 87 fountain clock and a rubidium 85 fountain clock.